\documentclass[12pt]{article}
\usepackage{amssymb,amsmath,graphicx,latexsym}\pdfoutput=1
\begin{document}
\title{\textbf{Monopoles, strings and dark matter}} 
\author{Catalina Gomez Sanchez\thanks{cgomez@physics.utoronto.ca}\hspace{2ex}and Bob Holdom\thanks{bob.holdom@utoronto.ca}\\
\emph{\small Department of Physics, University of Toronto}\\[-1ex]
\emph{\small Toronto ON Canada M5S1A7}}
\date{}
\maketitle
\begin{abstract}
We develop a scenario whereby monopoles in a hidden sector yield a decaying dark matter candidate of interest for the PAMELA and FERMI $e^\pm$ excesses. The monopoles are not completely hidden due to a very small kinetic mixing and a hidden photon mass. The latter also causes the monopoles and anti-monopoles to be connected by strings. The resulting long-lived objects eventually decay to hidden photons which tend to escape galactic cores before decaying. The mass scales are those of the hidden photon ($\approx 500$ MeV), the monopole ($\approx 3$ TeV) and the mixing scale (close to the Planck scale). A gauge coupling in the hidden sector is the only other parameter. This coupling must be strong and this results in light point-like monopoles and light thin strings.
\end{abstract}

\section{Introduction}
We shall describe a decaying dark matter scenario where the dark matter ``particle'' is a monopole and an anti-monopole connected by one or more strings. We shall refer to these objects as $MS\overline{M}$'s. Both the monopoles and the strings are composed of hidden sector fields. A nonabelian gauge symmetry of the hidden sector breaks to a $U(1)_h$ to produce monopoles and then the $U(1)_h$ breaks at a lower scale to produce strings. If the only remaining long range field to which the monopole couples is gravity then the $MS\overline{M}$'s can have cosmologically interesting life-times \cite{olum}. They can be considered for dark matter since their number densities are not constrained by the Parker bound \cite{parker}. The dynamics and evolution of monopoles attached to strings in the early universe has been quite well studied \cite{vilenkin,vil1,vil2,olum1,vash1}.

For the natural abundance of monopoles to be appropriate for dark matter they must be much lighter than standard GUT monopoles. This mass can in fact be in a range of interest for a decaying dark matter interpretation of the PAMELA \cite{pamela} and FERMI \cite{fermi} $e^\pm$ excesses. A $MS\overline{M}$ survives until the $M$ and $\overline{M}$ finally annihilate into hidden photons. If the hidden photon $\gamma_h$ experiences kinetic mixing \cite{holdom} with the photon and is otherwise stable then it decays into pairs of normal charged particles. With the appropriate mass its stable decay products are electrons, positrons and neutrinos \cite{nima, pospelov}.

We shall show in the next section that the kinetic mixing in combination with the hidden photon mass implies that a $MS\overline{M}$ will pick up the normal magnetic field of a dipole. These oscillating dipoles lose energy through normal electromagnetic radiation, and when the mixing parameter is extremely small the lifetime of the $MS\overline{M}$'s can be appropriate for decaying dark matter. Thus in our scenario the kinetic mixing is setting the lifetime for both the $MS\overline{M}$'s and the $\gamma_h$'s, and it is responsible for producing an observable signal.

We first summarize the various parameters and relations between them \cite{vilshell}. When the hidden nonabelian gauge symmetry breaks to $U(1)_h$ some gauge bosons receive mass $m_X$. If the gauge coupling is $e_h$ then the monopoles have mass
\begin{equation}
m_M\approx\frac{4\pi}{e_h^2}m_X
\end{equation}
and a size of order $m_X^{-1}$. We assume that the monopoles form at a temperature $T_M\approx m_X/e_h\approx m_M/g_h$ where $g_h=4\pi/e_h$ is the magnetic coupling. When the surviving $U(1)_h$ gauge symmetry breaks at a lower scale this hidden photon $\gamma_h$ develops a mass $m_h$. At this scale the coupling may have run to a new value $e_h'$. In the results to follow either $e_h$ or $e_h'$ should appear depending on the context, but for simplicity we shall drop the distinction and simply use $e_h$. Strings have an energy per unit length
\begin{equation}
\mu\approx\frac{\pi}{e_h^2}m_h^2
\label{e16}\end{equation}
and a thickness of order $m_h^{-1}$. We assume that the strings form at a temperature $T_S\approx m_h/e_h$.

The two mass scales of the hidden sector, $m_M$ and $m_h$, are fairly well determined if we are to make contact with the dark matter interpretations of PAMELA and FERMI data. For the stable products of $\gamma_h$ decays to be electrons, positrons and neutrinos only, $m_h$ could be anywhere from above the $e^+e^-$ threshold up to about a GeV. But a mass above the $\mu^+\mu^-$ and $\pi^+\pi^-$ thresholds is preferred since it gives a broader electron/positron spectrum to fit the FERMI data \cite{meade}. Given this and with the mass of the decaying $MS\overline{M}$ close to $2m_M$, the PAMELA data favors a $m_M$ in the 1 to 3 TeV range while the FERMI data favors a mass at the upper end of this range \cite{meade}. We shall adopt the values $m_h=500$ MeV and $m_M=3$ TeV for illustration. Adjustments in these masses are still possible.

The remaining parameters are the hidden sector gauge coupling $e_h$ and the kinetic mixing parameter $\chi$. In section 3 we show how the correct initial number density of monopoles constrains $e_h$. Here and at other points in our discussions we shall find that a strong coupling is required. The lifetime of the $MS\overline{M}$'s as determined by the emission of electromagnetic radiation must be appropriate for decaying dark matter. We study how this is possible in section 4 while in section 5 we consider other energy loss mechanisms. The $MS\overline{M}$'s that decay today are much smaller than average and we can determine enough about the distribution of these sizes so that we are able to fix $\chi$.  The lifetime of the $\gamma_h$ is then also determined. We find that this lifetime tends to be sufficiently long so that $\gamma_h$'s travel out of galactic cores before decaying, and this may have interesting consequences for the associated gamma ray signal. The lifetime of the mediator particle in secluded models of dark matter often face a constraint from big bang nucleosynthesis \cite{posp}. In section 6 we argue that in our case the $\gamma_h$'s are unstable in the presence of a light string network. In section 7 we briefly consider the late time properties of the dark matter and its self-interactions while in section 8 we look at the possibilities (or lack thereof) for its direct detection.

Here we can comment on the origin of the mixing parameter $\chi$. It is related to the mass scale of the physics responsible for the mixing between the hidden and standard model sectors. If this physics respects the hidden nonabelian gauge symmetry then the lowest dimensional operator that can give rise to the mixing is
\begin{equation}
\frac{1}{M_{mix}}Y_{\mu\nu}G^{\mu\nu}_{a}\phi_a
\label{e7}.\end{equation}
This couples the gauge bosons of hypercharge and the hidden gauge group to $\phi_a$, a hidden adjoint scalar field. If $\langle\phi_a\rangle$ signals the breakdown of the gauge symmetry then $\langle\phi_a\rangle\approx m_X/e_h\approx m_M/g_h$.  The induced kinetic mixing $\chi$ between the photon and the $\gamma_h$, as defined in the next section, is $\chi\approx\langle\phi_a\rangle/M_{mix}$ and so
\begin{equation}
M_{mix}\approx\frac{m_M}{\chi g_h}
\label{e6}.\end{equation}
(The absence of a field $\phi_a$ would mean that a higher dimensional operator is necessary which would imply a smaller mixing scale.) It would be appealing for $M_{mix}$ to be close to the Planck scale to avoid the introduction of another mass scale. Due to the very small value of $\chi$ required to produce a suitable $MS\overline{M}$ lifetime we shall find that $M_{mix}$ as given by (\ref{e6}) must indeed be of this size.

The strong coupling value that we have said is required for $e_h$ could also be viewed as natural, since then the symmetry breakings that are necessary in the hidden sector could be occurring dynamically. In this way are are encouraged to find interesting results where the parameters $e_h$ and $\chi$ are close to their ``natural'' values.

\section{Mixing and monopoles}
At low energies the electromagnetic $U(1)$ and the massive $U(1)_h$ gauge fields and their mixing are described by 
\begin{equation}
 \mathcal{L}=-\frac{1}{4}F'_{\mu\nu}{F'}^{\mu\nu}-\frac{1}{4}F'_{h\mu\nu}{F'}_h^{\mu\nu}-\frac{1}{2}\chi F'_{\mu\nu}{F'}_h^{\mu\nu}+\frac{1}{2}m_h^2A'_{h\mu}{A'}_h^{\mu}
.\label{e12}\end{equation}
Diagonal kinetic terms can be regained while retaining the masslessness of the photon by redefining the fields in terms of new fields $A$ and $A_h$ as
\begin{eqnarray}
A'_h &\rightarrow& A_h \nonumber\\
A' &\rightarrow& A-\chi A_h
.\label{e13}\end{eqnarray}
This means that all fields coupling to the photon with charge $e$ will pick up a coupling to the hidden photon of strength $-\chi e$ \cite{holdom}.

The $U(1)_h$ has emerged from the breakdown of a larger gauge group such that monopoles arise as regular solutions of the field equations. But the (hidden) magnetic charge of these monopoles must be quantized according to Dirac's quantization condition, and so there is a question of how this is compatible with particles with a hidden charge of $-\chi e$. This is answered in \cite{jaeckel} where it is shown that monopoles, including 't Hooft-Polyakov monopoles, can carry a combination of both magnetic charges. The argument for two massless $U(1)$'s is reproduced here, but adapted to our $A$ and $A_h$ basis.
 
If a charge $e$ is in the presence of a magnetic monopole with magnetic charge $g$ the angular momentum of the fields is
\begin{equation}
\boldsymbol{L}=\int d^3x\boldsymbol{x}\times(\boldsymbol{E}\times\boldsymbol{B})=\frac{eg}{4\pi}\hat{n}
.\end{equation}
This must be quantized
\begin{equation}
\vert \boldsymbol{L}\vert=\frac{eg}{4\pi}=\frac{n}{2},
\end{equation}
where $n$ is a non-negative integer. In the case of two $U(1)$'s the magnetic field of a monopole is
\begin{equation}
\begin{pmatrix}
\boldsymbol{B}\\
\boldsymbol{B_h}
\end{pmatrix}=\frac{\boldsymbol{r}}{4\pi r^3}\begin{pmatrix}
			g\\
			g_h\\
			\end{pmatrix},
\end{equation}
while the electric fields of normal and hidden charges are respectively
\begin{equation}
\begin{pmatrix}
\boldsymbol{E}\\
\boldsymbol{E_h}
\end{pmatrix}=\frac{e\boldsymbol{r}}{4\pi r^3}
			\begin{pmatrix}
			1\\
			-\chi
			\end{pmatrix},
\label{e21}\end{equation}
\begin{equation}
\begin{pmatrix}
\boldsymbol{E}\\
\boldsymbol{E_h}
\end{pmatrix}=\frac{e_h\boldsymbol{r}}{4\pi r^3}
			\begin{pmatrix}
			0\\
			1
			\end{pmatrix}.
\end{equation}
A system consisting of an ordinary charge and a monopole gives
 \begin{equation}
\vert\boldsymbol{L}\vert=\frac{-\chi e g_h+eg}{4\pi}=\frac{n}{2},
\end{equation}
while for a hidden charge and a monopole we have
\begin{equation}
\vert\boldsymbol{L'}\vert=\frac{e_h g_h}{4\pi}=\frac{m}{2}.
\end{equation}
These two equations give two types of allowed monopoles
\begin{equation}
M_{GUT} \rightarrow\begin{pmatrix}
g\\
g_h
\end{pmatrix}=\frac{2\pi n}{e}
			\begin{pmatrix}
			1\\
			0
			\end{pmatrix},
\end{equation}
and
\begin{equation}
M \rightarrow \begin{pmatrix}
g\\
g_h
\end{pmatrix}=\frac{2\pi m}{e_h}
			\begin{pmatrix}
			\chi\\
			1
			\end{pmatrix}.
\label{e14}\end{equation}
We shall comment on our choice $g_h=4\pi/e_h$ ($m=2$) in section 4.

We see that the combination of magnetic charges carried by the monopole $M$ is orthogonal to the combination of charges carried by a normal charge. As long as $\gamma_h$ remains massless these monopoles remain hidden to normal charges. However once $U(1)_h$ breaks, this hidden component of the $M$ field becomes confined for length scales larger than $m_h^{-1}$. Flux tubes, or strings, of the hidden magnetic field can begin on $M$'s and end on $\overline{M}$'s. Thus after $U(1)_h$ breaks the remaining long range field of $M$ is purely electromagnetic (the $\chi$ component in (\ref{e14})) and the hidden monopoles become visible to normal charges. The $M$'s now display an apparent violation of the Dirac quantization condition, but this is allowed due to the attached physical string(s). These strings are also visible to normal charges (due to the $\chi$ component in (\ref{e21})) through the Aharonov-Bohm effect as we discuss in sections 5 and 8.

Let us return to the point that the $U(1)_Y$ of hypercharge is involved in the origin of the mixing, as in (\ref{e7}). In the mass eigenstate basis this can be seen as a mixing of both the photon and the $Z$ boson separately with the hidden photon. The $Z$ mass defines a basis to describe its mixing, and so by the same arguments as before particles charged under $U(1)_h$ acquire a small $Z$ charge while the hidden monopoles do not acquire a $Z$ magnetic charge. In this way see that the hidden monopoles are not affected by electroweak symmetry breaking.

\section{Monopole densities}
The hidden monopole $M$ number density is $n_M$ and it is equal to the $\overline{M}$ density at all times. For a $M$ mass of 3 TeV we wish to investigate the conditions under which $n_M$ could be appropriate for dark matter. We first consider the case when the magnetic charge is large as in the case of GUT monopoles, so that we can apply the same analysis \cite{preskill}. (This is the weak gauge coupling case and we turn later to strong coupling.) The density of monopoles when they are first formed at temperature $T_M$ can be reduced through annihilation of $M$-$\overline{M}$ pairs. This occurs if there are light particles carrying the hidden charge in the plasma. Such fields should be present in our scenario since they are needed at a much lower energy scale to produce an order parameter for the breaking of $U(1)_h$. A $M$ drifting through such a plasma towards an $\overline{M}$ can experience energy loss, capture and thus annihilation. When the mean free path becomes longer than the capture distance the process ends at some temperature $T_f$ \cite{preskill},
\begin{equation}
T_f\approx\frac{m_M}{B^2}\frac{1}{\alpha_M^2},\quad\quad B=\frac{3\zeta(3)}{4\pi^2}\sum_i (\frac{g_he_h^i}{4\pi})^2
.\end{equation}
$\alpha_M\equiv g_h^2/4\pi$ and the sum is over all spin states of relativistic hidden charged particles. The condition $T_f<T_M\approx m_M/g_h$ implies $\alpha_M\gtrsim(4\pi/B^4)^{1/3}$. If this is satisfied then an initial value of $n_M/T^3$ larger than the following will be reduced to the following
\begin{equation}
\frac{n_M}{T^3}\approx\frac{\kappa^3 m_M}{4\pi BCm_{Pl}}\frac{1}{\alpha_M^3},\quad\quad C=\left(\frac{45}{4\pi^3g_*}\right)^\frac{1}{2}
\label{e1}
.\end{equation}
The $\kappa$ factor is introduced since a temperature such as $T_f$ is a hidden sector temperature, and this may be a factor of $\kappa$ times the temperature $T$ of the observable sector. $g_*$ is the usual total effective number of relativistic spin degrees of freedom. For the smallest $\alpha_M$ at which this annihilation process is still operative the remaining monopole abundance would be 3 or 4 orders of magnitude below what is required for dark matter (for $C=.05$ and $\kappa=1$). Thus $\alpha_M$ must be smaller to turn off the annihilations ($T_f>T_M$), and this puts a lower bound on $\alpha_h=1/\alpha_M\gtrsim 1/2$ for $B\approx1$.

We thus turn to the initial value of $n_M/T^3$. If we still believe that $\alpha_h$ could be fairly weak then the density of monopoles produced in a second order phase transition can be related to the correlation length $\xi$ and the relaxation time $\tau$ as the system passes through the phase transition at a finite speed. The speed is determined by the Hubble parameter $H$ at that time. This is the Kibble-Zurek mechanism \cite{kibble,zurek} (reviewed in \cite{murayama}). In terms of critical exponents defined from
\begin{eqnarray}
\xi&=&\xi_0\vert\varepsilon\vert^{-\nu}\\
\tau&=&\tau_0\vert\varepsilon\vert^{-\mu}
\end{eqnarray}
where $\varepsilon=(T_M-T)/T_M$, the following result is obtained
\begin{equation}
\xi\approx\left(\frac{\sqrt{\lambda}T_M}{H}\right)^{\frac{\nu}{1+\mu}}\frac{1}{\sqrt{\lambda} T_M}
.\label{e5}\end{equation}
$\lambda$ is the coupling appearing in a Ginzburg-Landau approximation to the free energy. Since $n_M\approx\xi^{-3}$ this leads to
\begin{equation}
\frac{n_M}{T^3}\approx \kappa^3\lambda^{3/2}\bigg(\frac{1}{\kappa^2\lambda^{1/2}C}\frac{m_M}{g_h m_{Pl}}\bigg)^{\frac{3\nu}{1+\mu}}
.\label{e3}\end{equation}
The classical values of the exponents are $\nu=\mu=1/2$ which makes the exponent in (\ref{e3}) unity. With $\lambda=\kappa=1$ and using the upper bound on $\alpha_M=g_h^2/4\pi$ from above, the monopole density would be about two orders of magnitude too small. This could be corrected if $\mu$ and $\nu$ deviated from their classical values in such a way as to reduce the exponent in (\ref{e3}) (the causality constraint is $\nu\le\mu$). Otherwise we are pushed towards still smaller $\alpha_M$ (larger $\alpha_h$) and larger $\lambda$. In fact the Kibble-Zurek mechanism becomes irrelevant when $\alpha_h$ becomes larger than unity.

When $\alpha_h>1$ the monopoles become lighter than the massive gauge bosons, $m_M\approx m_X/\alpha_h$. The monopole has a size $\approx1/m_X$ which is smaller than its Compton wavelength. These relatively light point-like monopoles can be treated like normal particles. With their fairly weak magnetic charge they will experience pair production and annihilation through two to two processes, and in this way they will reach thermal equilibrium with the light degrees of freedom in the dark sector. The monopoles will remain in thermal equilibrium until the temperature falls sufficiently below $m_M$. The final freeze-out temperature is reached when $n_M\langle\sigma v\rangle\approx H$ where $\langle\sigma v\rangle$ is the annihilation cross section. We assume that this annihilation is analogous to two charged scalars annihilating into 2 photons \cite{pospelov} so that
\begin{equation}
\langle\sigma v\rangle=\frac{\pi\alpha_M^2}{m_M^2}
.\end{equation}
This has to be close to the usual value of $3\times10^{-26}$ cm$^3$s$^{-1}$ to arrive at the correct dark matter abundance \cite{kolb}. After including the dependence on $\kappa$ we obtain
\begin{equation}
\alpha_M\approx\frac{1}{4\pi}\sqrt{\kappa}\left(\frac{m_M}{\mbox{2.8 TeV}}\right)
\label{e10}.\end{equation}
In other words $\alpha_h=1/\alpha_M\sim4\pi$ and we note that this agrees with the definition of strong coupling in ``naive dimensional analysis'' \cite{cohen}. It suggests that gauge symmetries are breaking dynamically in the hidden sector. Here we note that $e_h$ always represents the charge of the order parameter and so in the case that the latter is a fermion condensate the actual gauge coupling of the fermions is $\overline{e}_h=e_h/2$ and so $\overline{\alpha}_h\sim\pi$.

Thus with a monopole mass close to 3 TeV, as hinted at by PAMELA and FERMI data, the required monopole abundance for dark matter leads us to the strong coupling case. The relations between the various masses and couplings that we mentioned in the introduction are assumed to extrapolate into this regime. In some sense it is not a severe extrapolation, since when $\alpha_M\approx1/4\pi$ or $g_h\approx 1$ the monopole mass has only come down to the scale of symmetry breaking as given by $m_M/g_h\approx m_X/e_h$.

\section{Lifetimes}
Below some temperature the $U(1)_h$ breaks and the hidden photon develops a mass $m_h$. $M$-$\overline{M}$ pairs become connected by strings to form $MS\overline{M}$'s. Here we need to be a little more explicit about how the gauge symmetries are breaking to form monopoles and strings. Let us consider the simplest example, the breakdown of $SU(2)$ to $U(1)_h$ which then itself breaks. If the first step occurs via a scalar triplet $\langle\phi_a\rangle$ then $m=2$ in (\ref{e14}) as we have been assuming. A second scalar triplet with a smaller vacuum expectation value can be used to break $U(1)_h$. In this case the flux carried by a string will be $2\pi/e_h$ and this implies that each monopole will end up with two strings attached. (We assume that two $2\pi/e_h$ strings are energetically favored over a single $4\pi/e_h$ flux string.) If instead a scalar doublet is used to break the $U(1)_h$ then there are only $4\pi/e_h$ strings and each monopole will have one string attached. 

In the first case a ``necklace'' can also form where equal numbers of monopoles and anti-monopoles are attached to one loop of string. The evolution of necklaces in the early universe was studied in \cite{olum1} where it was concluded that necklaces tend to cut themselves up into a set of $MS\overline{M}$'s and pure string loops. The evolution of an isolated $MS\overline{M}$ is similar whether it has one or two strings and so the results in the two cases will be similar. On the other hand we have mentioned dynamical symmetry breaking where it is natural to consider a condensate of fermions rather than a vev of a scalar field. Weyl fermions that transform as doublets under $SU(2)$ could develop Majorana condensates that transform as triplets under the $SU(2)$. This suggests that the triplets only case (the two string case) may be more natural, and we shall assume this in the following.

The acceleration of the $M$ caused by the two strings, assuming they are pulling in the same direction, is $a=2\mu/m_M$. The strings in the $MS\overline{M}$'s should be fairly straight for various reasons. Strings can be fairly straight on formation, especially for the smaller $MS\overline{M}$'s. In the next section we look at mechanisms by which strings very slowly lose energy. But most importantly, a $MS\overline{M}$ with an excess amount of string can emit a loop of string, since when a piece of string intersects itself it may pinch off to form a loop. Thus the $MS\overline{M}$ should end up in a state where the strings remain quite straight as the $M$ and $\overline{M}$ move around their center of mass, with energy moving back and forth between monopole kinetic energy and string rest mass energy. Any angular momentum of the system will keep the $M$ and $\overline{M}$ from colliding. We will consider the peak velocities of the monopoles below, but they can be substantially larger than the typical monopole velocities at the time of string formation.

We will need to determine a distribution of lifetimes of the collection of $MS\overline{M}$'s after they have started to evolve as isolated systems. Some will have already decayed by now, but this is a very small fraction of the original number as is usual with decaying dark matter. We define the probability that a randomly chosen $MS\overline{M}$ will decay between time $t$ and $t+dt$ as $P(t,\overline{\tau})dt$, where $\overline{\tau}$ is the mean lifetime. Then the rate of decay per unit volume at the present time $t_0=1/H$ is $n_0 P(t_0,\overline{\tau})$ where $n_0$ is the present density. We can write $P(t_0,\overline{\tau})=\tau_{\rm eff}^{-1}$ where $\tau_{\rm eff}$ is an effective lifetime, to be distinguished from the actual mean lifetime $\overline{\tau}$ of the $MS\overline{M}$'s.

For now we focus on the energy loss due to normal electromagnetic radiation; in the next section we shall compare this to other possible energy loss mechanisms. The Larmor formula for a charge with proper acceleration $a$ can be used to find the power radiated by the $M$ and $\overline{M}$ even in the relativistic case \cite{vil2},
\begin{eqnarray}
\frac{d\mathcal{E}_{em}}{dt}&=&-2\frac{\chi^2}{6\pi}(g_ha)^2= -\frac{\chi^2}{3\pi}\bigg(\frac{4\pi}{e_h}\frac{2\mu}{m_M}\bigg)^2\nonumber\\
  					    &\approx& -\frac{64\pi^2}{3}\frac{m_h^2}{e_h^4m_M^2}\chi^2\mu
. \label{e19}\end{eqnarray}
Our choice of $\chi$ will make this very small, but it is still larger than the power radiated into the massive $\gamma_h$. The latter is exponentially suppressed \cite{vil2} with a factor $\exp(-2vm_h/3a)\sim\exp(-4v\alpha_h m_M/3m_h)$. For fairly straight strings the total energy to lose is $\approx2\mu L$ where $L$ is the maximum separation of the $M$ and $\overline{M}$. Then (\ref{e19}) gives a lifetime that is simply proportional to $L$
 \begin{equation}
\tau_{em}\approx\frac{3}{2}\alpha_h^2\frac{m_M^2}{m_h^2}\frac{L}{\chi^2}
.\label{e4} \end{equation}
Thus the distribution of lifetimes in the collection of $MS\overline{M}$'s is determined by the distribution of $L$'s.

In this context it is useful to consider the distribution of nearest neighbor distances for a random set of points in 3 dimensions. The nearest neighbor distribution can be derived from the relation
\begin{equation}
P_{nn}( r,n) =4\pi n{r}^{2}\left( 1-\int_{0}^{r}P_{nn}( s,n) ds\right) 
\end{equation}
where $n$ is the density of points. By solving this relation for $P_{nn}( r,n)$ one finds that the mean $r$ of the nearest neighbor distribution is
\begin{equation}
\overline{r}_{nn}=\frac{\gamma(\frac{1}{3})}{(36\pi n)^{1/3}}\approx0.554n^{-1/3}
\end{equation}
where $\gamma(x)$ is the gamma function. Eliminating $n$ in favor of $\overline{r}_{nn}$ gives
\begin{equation}
P_{nn}(r,\overline{r}_{nn})=\frac{{\gamma( \frac{1}{3}) }^{3}{r}^{2}}{9\overline{r}_{nn}^{3}}\exp\left(-\frac{{\gamma( \frac{1}{3}) }^{3}{r}^{3}}{27\overline{r}_{nn}^{3}}\right)
.\label{e20}\end{equation}
This satisfies $\int_0^\infty P_{nn}(r,\overline{r}_{nn})dr=1$ and $\int_0^\infty rP_{nn}(r,\overline{r}_{nn})dr=\overline{r}_{nn}$.
The lifetime is proportional to $L$, and if we associate $L$ with the nearest neighbor distance $r$ then we can obtain a distribution of  ``nearest neighbor lifetimes'',
\begin{equation}
P_{nn}(t,\overline{\tau}_{nn})=P_{nn}(r,\overline{r}_{nn})|_{(r\rightarrow t,\;\;\overline{r}_{nn}\rightarrow\overline{\tau}_{nn})}
.\end{equation}
$P_{nn}(t,\overline{\tau}_{nn})$ is not a physical distribution of lifetimes since it is not even possible for every $M$ to be connected to its nearest $\overline{M}$ and vice versa.

But we wish to argue that $P_{nn}(t,\overline{\tau}_{nn})$ for some $\overline{\tau}_{nn}$ is a good approximation to $P(t,\overline{\tau})$ when $t\ll\overline{\tau}_{nn}$. This corresponds to $M$-$\overline{M}$ pairs with separations that are much smaller than $\overline{r}_{nn}$, and it is precisely for these pairs that it is very likely that they are connected by strings. Thus for these very close pairs the distribution of $L$'s should be quite similar to the distribution of nearest neighbor separations. These small $MS\overline{M}$'s include those that are decaying today and so we can determine $\overline{\tau}_{nn}$ by setting $P_{nn}(t_0,\overline{\tau}_{nn})=\tau_{\rm eff}^{-1}$ where $t_0=1/H$. A typical value of $\tau_{\rm eff}$ for decaying dark matter is $2\times10^{26}$ sec, and from this we obtain $t_0/\overline{\tau}_{nn}\approx 10^{-3}$. We also have the relation $L_0/\overline{r}_{nn}=t_0/\overline{\tau}_{nn}$. $L_0\approx10^{-3}\overline{r}_{nn}=0.554\times10^{-3}n_M^{-1/3}$ is the initial size of those $MS\overline{M}$'s that are decaying today. In the following sections it will become clear that $P_{nn}(t,\overline{\tau}_{nn})$ will be a poor approximation to $P(t,\overline{\tau})$ for times far in the future $t\gg t_0$.

$n_M^{-1/3}$ is obtained by scaling the present value of $n_0^{-1/3}$ for dark matter back to its value at string formation when the temperature is $\approx m_h/\kappa e_h$. The present temperature $T_0$ is enhanced by 1.4 due to the annihilation of $e^+e^-$ at an intermediate temperature and so
\begin{equation}
L_0\approx0.55\times10^{-3}\frac{\kappa e_h}{m_h}\frac{T_0}{1.4}\left(\frac{\rho_0}{2m_M}\right)^{-1/3}
\label{e9}.\end{equation}
Note that the explicit $\kappa$ dependence will cancel the $\kappa$ dependence of $\alpha_h^2$ in (\ref{e4}) due to (\ref{e10}), and henceforth we set $\kappa=1$. The resulting $L_0\approx 2\times10^{-12}$ cm is small in the sense that it is only about 50 times larger than the thickness of the string $\approx 1/m_h$. For these small $MS\overline{M}$'s the resulting peak velocities of the monopoles are
\begin{equation}
v\approx\bigg(\frac{2\mu L_0}{m_M}\bigg)^{1/2}\approx0.02
.\label{e8}\end{equation}
Much larger $MS\overline{M}$'s can have relativistic internal motions.

We can now determine $\chi$ by setting $\tau_{em}$ from (\ref{e4}) to $t_0=1/H$ after replacing $L$ with $L_0$. We obtain $\chi\approx 1.2\times10^{-15}$. From (\ref{e6}) we find that the scale of physics responsible for the mixing can be as high as $M_{mix}\approx 3\times10^{18}$ GeV $\sim m_{Pl}$ as advertised.

When the $MS\overline{M}$ has lost sufficient energy so that the separation of the $M$-$\overline{M}$ pair remains less than the string thickness, then the string dynamics no longer plays a role. The $M$-$\overline{M}$ pair forms a fairly weakly bound ``monopolonium'' state \cite{hill}. It cascades down to the $n=1$ ground state which has binding energy $R=m_M/4\alpha_h^2 $. The classical lifetime for a starting radius of $m_h^{-1}$ is $\tau=\alpha_h^2 m_M^2/8m_h^3\chi^2$. This is much shorter than the original $MS\overline{M}$ lifetime, and it may be even shorter still due to the emission of $\gamma_h$'s through quantum $\Delta n>1$ transitions. $\Delta n=1$ transitions involving $\gamma_h$'s are only possible for the lowest levels, $n<(2R/m_{\gamma_h})^{1/3}\sim3$. Only a few of these low energy $\gamma_h$'s are produced since $R/m_h\approx10$. Once in the $n=1$ state the $M$-$\overline{M}$ pair finally annihilates to $2\gamma_h$ (usually only two due to the fairly small magnetic coupling).

These final $\gamma_h$'s are highly relativistic and since their coupling to charged matter is so small they can travel a long distance $d_h$ before decaying,
\begin{equation}
d_h=\gamma\tau_hc\approx\frac{m_M}{m_h}\frac{2}{\alpha\chi^2m_h}c
\label{e18}.\end{equation}
From the values of parameters as already given this is about 15 kpc. $d_h$ of this order implies that the $\gamma_h$'s will tend to decay away from the regions where the dark matter densities are the highest, such as galactic cores. Also the $\gamma_h$'s that give rise to the observed $e^\pm$'s will originate from more distant parts of the galactic dark matter halo, but for $d_h\approx15$ kpc the $e^\pm$ flux is only reduced by about 30\%. On the other hand $d_h$ cannot be much larger than this. Since $d_h\propto e_h^{-5}$, due to how $\chi$ is determined by (\ref{e4}) and (\ref{e9}), we see this as another constraint that rules out small coupling.

If most $\gamma_h$'s originating from our galactic core have not decayed by the time they reach earth then this also affects the associated gamma ray signal \cite{rothstein}. In particular this signal should show less enhancement in the direction of the galactic core and thus be more isotropic than expected \cite{ibarra}.  The dominant part of the gamma ray signal arises from the $e^\pm$'s up-scattering background photons. The inter-stellar radiation field has a harder spectrum in the galactic core, and these photons when up-scattered produce higher energy gamma rays. Thus the typical energy of the gamma rays is also reduced when the $\gamma_h$'s decay outside the galactic core. The same effect applies to gamma ray signals from decaying dark matter in nearby clusters of galaxies \cite{profumo}; the typical gamma ray energies are reduced. The gamma ray signal that is not affected comes from the up-scattering of CMB photons and the observational constraint on this signal is the same as in other decaying dark matter models.\footnote{This constraint keeps one from arbitrarily decreasing $\tau_{\rm eff}$ to compensate for a larger $d_h$, since decreasing $\tau_{\rm eff}$ would increase the CMB gamma ray signal as well as the $e^\pm$ signal.}

\section{More energy loss}
The moving $M$ and $\overline{M}$ emit gravitational radiation, and the energy loss rate as estimated in \cite{olum} (with $\mu$ replaced by $2\mu$) is\footnote{The relativistic case is studied in \cite{martin}.}
\begin{equation}
\frac{d\mathcal{E}_g}{dt}\approx-\frac{256}{5}\frac{G\mu^3 L}{m_M}.
\end{equation}
For the $MS\overline{M}$'s that decay today this is insignificant compared to the electromagnetic radiation studied in the last section. But far enough in the future and for those $MS\overline{M}$'s that still survive this energy loss can become dominant. Then $L(t)= L_i e^{-t/\tau_g}$ where
\begin{equation}
 \tau_g\approx\frac{5}{256}\frac{m_M}{G\mu^2}.
 \end{equation}
This is about 400 times the age of the universe and up to a logarithmic dependence on $L_i$, this sets the maximum lifetime of any $MS\overline{M}$. In particular the lifetime is no longer proportional to $L$ on these time scales.

A string loop emits gravitational radiation at a rate \cite{casper}
\begin{equation}
\frac{d\mathcal{E}_g^s}{dt}\approx-50G\mu^2
,\label{e17}\end{equation}
and so a loop with length $\pi L$ has a lifetime
\begin{equation}
\tau_g^s\approx \frac{\pi L}{50G\mu}
.\end{equation}
Thus a string loop will decay within the age of the universe if $L\lesssim10^{-11}$ cm. The lifetime of a loop of length $\pi L$ turns out to be similar to the lifetime of a $MS\overline{M}$ with size $L$, since $\tau_g^s/\tau_{em}\approx1/5$. The gravitational radiation by the strings in a $MS\overline{M}$ has the effect of causing these strings to lose excess kinetic energy and to straighten.

We now turn to frictional effects \cite{vilshell}. The normal magnetic fields of the $M$ and $\overline{M}$ will induce frictional effects through interactions with charged particles in the plasma,
\begin{equation}
\frac{d\mathcal{E}_f}{dt}\approx-\chi^2B'T^2v^2
\end{equation}
where $B'\approx\pi^{-2}\sum_i (e_ig_h/4\pi)^2$. For friction to remove the energy $2\mu L$ and for the $v$ in (\ref{e8}) we obtain
\begin{equation}
\tau_f\approx\frac{2\mu L}{\chi^2B' T^2 v^2}\approx\frac{m_M}{\chi^2B'T^2}
.\end{equation}
But due to the tiny $\chi$ it is easy to see that $m_M/\chi^2B'\gg m_{Pl}$ and thus $\tau_f$ is much larger than the Hubble time. This friction can be neglected.

Normal charged particles can also scatter off a string via an Aharonov-Bohm (AB) effect, since charged particles carry an hidden charge $\epsilon e_h\equiv-\chi e$ while the strings carry a $2\pi/e_h$ unit of hidden flux. The resulting AB cross section per unit length is \cite{alford,vilshell}
\begin{equation}
\frac{d\sigma_{AB}}{dl}=\frac{2}{p}\sin^2(\pi\epsilon)
\end{equation}
where $p$ is the transverse particle momentum. One can use this to estimate the time on which the transverse velocity of a section of string will be damped out as
\begin{equation}
\tau_{AB}\approx\frac{\mu}{\epsilon^2B'' T^3}
.\end{equation}
$B''\approx 2\sum_a b_a$ where $b_a$ is $3/4$ for fermions and $1$ for bosons. Here again $\mu/\epsilon^2B''T\gg m_{Pl}$ and so this effect can be neglected.

Since the hidden magnetic fields of the monopoles have been confined to strings carrying a unit of hidden flux, particles of the hidden sector that carry a unit of hidden charge will not contribute to the two previous frictional mechanisms. On the other hand a particle with hidden charge $e_h/2$ will have AB scattering at full strength, $d\sigma_{AB}/dl=2/p$. This can be the case for the fermions that develop the $U(1)_h$ breaking condensate. Their masses should be similar to $m_h$, and in fact they must be greater than $m_h/2$ to prevent $\gamma_h$ from decaying into them.\footnote{There need not be much lighter or massless particles remaining in the hidden sector. If there are we would need to assume that the $\gamma_h$ does not decay into them, otherwise the decays of $MS\overline{M}$'s are undetectable. But then it is unlikely that such particles will interact with the string and so they are irrelevant for the string dynamics.}  There can also be a direct interaction of hidden particles with the fields of the string, which for small $p/m_h$ has a cross section \cite{everett}
\begin{equation}
\frac{d\sigma_{\rm dir}}{dl}\approx\frac{\pi^2}{p\ln(p/m_h)^2}
.\label{e15}\end{equation}
Thus there could be a damping of transverse motions of the strings if they move in a bath of the massive particles with which they interact with cross sections $\sigma_{AB}$ and/or $\sigma_{\rm dir}$. In the next section we discuss why there is very little such damping.

The Aharonov-Bohm effect implies that the string interacts with electrons through the vector electron current. This gives rise to a more interesting process where the string emits $e^+e^-$ pairs. For a fairly smooth string loop of length $\pi L$ the rate of energy loss estimated on dimensional grounds is \cite{vash1}
\begin{equation}
\frac{d\mathcal{E}_{ee}^s}{dt}\approx-\frac{\epsilon^2}{L^2}\quad\mbox{for}\quad L\lesssim\frac{1}{2m_e}
,\label{e22}\end{equation}
which in turn gives a lifetime
\begin{equation}
\tau_{ee}^s\approx\frac{\pi\alpha_h\mu L^3}{3\alpha\chi^2}
.\label{e23}\end{equation}
For $L\approx1/2m_e\approx2\times10^{-11}$ cm the lifetime is $\tau_{ee}^s\approx10^8$ years. For larger $L$ the process is exponentially suppressed, so that loops more than about 3 times as large can last longer than the age of the universe. In any case we see that this process removes the smaller loops at least as effectively as gravitational radiation.

$e^+e^-$ emission can be enhanced if the string loop has kinks and/or cusps. Cusps tend to form on featureless loops with little excitation of higher string harmonics, but such loops are less likely when there is little damping. Kinks also inhibit cusps \cite{garf} and since kinks readily form when strings intercommute, they are expected to dominate. Kinky loops of any size could emit $e^+e^-$ pairs to produce an energy loss which is optimistically of order
\begin{equation}
\frac{d\mathcal{E}_{ee}^s}{dt}\approx-\frac{\epsilon^2}{L^2}\sqrt{\mu}\min(L,\frac{1}{2m_e})
.\end{equation}
Here we have included the likely effect that the electron mass has on the results of \cite{vash4}, and the extra factor compared to (\ref{e22}) is at most a factor of 70. This is an optimistic estimate for the rate since back-reaction effects will tend to flatten out the  kinks (most of the radiation comes from the sharper kinks) and thereby reduce the rate. Thus the presence of kinks probably does not dramatically increase the number of loops that can decay. We also note that the kinematics of kinks on the strings of a $MS\overline{M}$ can be quite different; on a loop the kinks move at the speed of light while on a $MS\overline{M}$ a kink can move slowly or even be stationary (as in a triangular standing wave).

The AB electron-string interaction also implies that a photon-string interaction will be generated through an electron loop \cite{vash1}. A current involving photons that can be induced by the vector electron current and which doesn't vanish for on-shell photons will involve 3 photons. The lowest dimensional current that can arise after integrating out the electron must then involve three factors of the photon field strength and one extra derivative. It will therefore be suppressed by a $1/m_e^4$ factor. For a string loop of length $\pi L$, now with $L$ larger than $1/2m_e$, the rate will be roughly
\begin{equation}
\frac{d\mathcal{E}_{\gamma\gamma\gamma}^s}{dt}\approx-\frac{\alpha^3}{(2m_e)^8}\frac{\epsilon^2}{L^{10}}
.\end{equation}
So even though this rate is not exponentially suppressed for large $L$, it has the $\alpha^3$ suppression and it still drops very quickly for increasing $L$. It thus has negligible effect. There is also the production of photons through their gravitational coupling \cite{garriga,vash2,vash3}, but this is proportional to $(G\mu)^2$ instead of $\epsilon^2$ and is thus miniscule.

\section{Strings and $\gamma_h$'s}

The picture of a dilute network of strings interacting with a dense bath of particles is completely altered at strong coupling. The strong coupling has an effect on the string forming transition similar to its effect on the monopole forming transition. In the latter case we saw that the monopoles became point-like and light compared to the massive elementary degrees of freedom (the massive gauge fields and other massive matter). Thus the monopole abundance was not determined by the correlation length (the Kibble-Zurek mechanism) but rather by thermal equilibrium. In the string forming transition it is the string tension $\approx m_h^2/4\alpha_h^2$ that is small relative to particle masses, and so the effective theory is a theory of light, thin strings.\footnote{The strong interaction limit of Nielsen-Olesen strings was considered in the original reference \cite{nielsen}.} These strings also carry fermion zero modes \cite{jackiw}. Thus the strings tend to capture fermions and the $\gamma_h$'s effectively have a decay channel into these zero modes. For example a $\gamma_h$ could decay into a zero mode and a normal fermion by interacting with the string. All this indicates that the relevant degrees of freedom after the transition are the strings and their excitations.

Thus the energy that was in a plasma of massless $\gamma_h$'s and fermions before the transition will be mostly deposited into strings and their kinetic energy after the transition. Due to the energy available in this plasma before the transition, the initial density of strings will be much higher than what the correlation length suggests. The string-string interaction rate will be very high (a section of string of length characteristic of the string spacing will collide many times in the Hubble time) and this should keep most of the total length of string in contact with the rest of the network. Only the loops that are very small may effectively be decoupled soon after the strings form. This should also be the case for the $MS\overline{M}$'s that are small enough to decay today, as we have already discussed.

The dense network of strings persists since it is undamped and since the energy loss from radiation (gravitational and electromagnetic) is so slow. This is quite unlike the  more standard evolution of cosmic strings which involves significant damping after formation. Due to energy considerations the strings are expected to move relativistically $\langle v_s^2\rangle\approx1$. A network of strings has an effective pressure \cite{vil1}
\begin{equation}
p_s=\frac{1}{3}(2\langle v_s^2\rangle-1)\rho_s
,\end{equation}
and so relativistically moving strings behave like a normal relativistic gas with $p=\rho/3$. In this limit the energy density in the string network does not increase relative to the total energy density. As long as $\rho_s$ remains a fairly small fraction of the total $\rho$ down to temperatures of about 1 MeV then it avoids constraints from big bang nucleosynthesis. Strings only form at a temperature of order $m_h/e_h\approx40$ MeV and so this is not a severe constraint.\footnote{We also note that the BBN constraints have recently weakened \cite{bbn}, and may even be in line with indications of new relativistic degrees of freedom from CMB studies \cite{cmb}.}

For weak coupling, which we have already argued is not interesting for other reasons, the $\gamma_h$'s and massive fermions would remain after the transition. Then in particular the $\gamma_h$, with its 500 MeV mass and a lifetime longer than 1 second, would not satisfy the BBN constraints \cite{posp}.\footnote{If the $\gamma_h$'s annihilated fast enough into much lighter or massless particles of the hidden sector then presumably they would also decay into these same particles, which would make them undetectable. Alternatively a $\gamma_h$ mass closer to a MeV and/or a small $\kappa$ could be considered in the weak coupling case.}

 The violent motions of the strings and their collisions could in principle produce $\gamma_h$'s. But unlike the case of $e^+e^-$ emission, the mass $m_h$ here is larger than $\sqrt{\mu}$ and so we expect substantial exponential suppression. Any $\gamma_h$'s that are produced face the prospect of string catalyzed decay back into dark sector degrees of freedom. In the absence of strings the proper lifetime of the $\gamma_h$ is about 10 years, so at that time $\gamma_h$'s can decay to ordinary leptons. But we expect that $\gamma_h$ production and decay only transfers a minor amount of energy from the hidden sector to the observable sector.

\section{Late times}
As the universe expands the interaction rate between $MS\overline{M}$'s gradually decreases, both because their separation increases and because their momenta are being redshifted. Starting from the smallest ones, the size of the $MS\overline{M}$'s that have a collision time larger than the Hubble time gradually increases and eventually most $MS\overline{M}$'s stop colliding. Loops free of monopoles will also be present. Loops of complicated shapes will quickly fragment into simpler loops which no longer self-intersect \cite{press}. Thus in addition to the $MS\overline{M}$'s there will be a population of loops which also stop interacting and which extend down in size to where they can decay to $e^+e^-$ pairs and gravitons within the age of the universe.

The relativistic string network should also survive to late times. The typical separation between strings in this network is much smaller than the Hubble scale, but its growth is proportional to $t$. Loops, with or without monopoles, of size anywhere close to this typical separation would be continually interacting and reconnecting with the rest of the network. As this separation increases there is an increasing population of smaller loops and $MS\overline{M}$'s that are more or less decoupled from the network. What is left of the relativistic string network today cannot have an energy density larger than the CMB, and so the typical separation in the string network now can be no less than about $3\times10^{10}$ cm. This of course is much larger than the present separation between $MS\overline{M}$'s.

For the dark matter that accumulates in halos, the increased density and speeds of the $MS\overline{M}$'s and loops leads to a resurgence of their interactions. Let us consider a nontrivial interaction between two $MS\overline{M}$'s with fairly straight strings of length $L_1$ and $L_2$ respectively. The cross section is $\sigma=\zeta L_1L_2$ where $\zeta$ accounts for the relative orientations, the fact that each is oscillating, and the probability for intercommuting. We take $\zeta=0.1$. Then the cross section corresponding to $L_1=L_2=10^{-9}$ cm for example would result in a rate of collisions of $n\sigma v\approx50H$ for local dark matter densities and speeds. This is a value of interest for self-interacting dark matter models \cite{spergel} and it suggests that the average $L$ cannot be much larger than $10^{-9}$ cm. A new feature here is that the cross section depends on the $L_i$'s of the $MS\overline{M}$'s being scattered. The smaller $MS\overline{M}$'s have a lower collision rate and so there can be a nearly collisionless subpopulation of $MS\overline{M}$'s. Also the collisions are not elastic since the sum of the internal energies can change after a collision. The effect that these new features have on the dynamics of dark matter halos remains to be explored.

Also of interest is the probability for two $MS\overline{M}$'s to collide to form a small $MS\overline{M}$ with a size no larger than $L_0\approx2\times10^{-12}$ cm, so that it can decay within the Hubble time. For this to happen a $M$ and $\overline{M}$ must be close together not only in position space but also in momentum space, since the kinetic energies of the monopoles in their center of mass frame cannot be larger than the rest mass of a segment of string of length $L_0$. Also the strings must intercommute at a point which will produce the small segment. Given the large dimensional phase space of possibilities, the production of a small $MS\overline{M}$ is highly suppressed. In fact our previous estimate for the number of small $MS\overline{M}$'s is due to a phase space suppression as well (e.g. the small $r$ behavior of (\ref{e20})). There the suppression is not as strong since only position space was involved; the kinetic energies of the monopoles at the time of string formation could be neglected. Thus the rate of $MS\overline{M}$'s decaying today should not be substantially changed from our previous estimate.

Collisions among a population of loops and $MS\overline{M}$'s also provide some probability for producing loops a few times $1/2m_e\approx2\times10^{-11}$ cm in size or smaller. As described in the last section these loops decay mostly into $e^+e^-$ pairs. $MS\overline{M}$'s similarly small in size could also emit $e^+e^-$ pairs if their strings are excited. The energies of the $e^+e^-$ pairs are typically not far above threshold and there are up to $\sim10^3$ such pairs produced from the largest of the small loops which can decay. The production of these small loops need not be very efficient to be of interest for the 511 KeV photon flux observed to come from the galactic bulge \cite{jean}. For example a collision can excite the strings of two $MS\overline{M}$'s, and even when these $MS\overline{M}$'s are not small it is still possible that they can emit a loop that is small enough to decay. From the analysis of \cite{maxim} we estimate that this probability would have to be smaller than about $10^{-6}$, assuming an average $L$ of $10^{-9}$ cm.

Finally we comment on large $MS\overline{M}$'s. When $L\approx 1$ angstrom ($10^{-8}$ cm $\approx5000L_0$) the energy in the strings is comparable to the energy in the monopoles $\mu L\approx m_M$. For larger $L$ the $MS\overline{M}$'s have relativistic monopoles and a total energy larger than $2m_M$. But we don't expect the average $L$ to be this large due to the constraints on the self-interactions. We note in passing that for $\mu L\gg m_M$ the ultra-relativistic monopoles can result in the emission of particles with energies $\gamma a\approx (\mu L/m_M)(2\mu/m_M)$ \cite{vil2}. For $\gamma_h$'s to be produced this energy must be greater than $m_h$, and this would require a very large $L>(m_M/\mu)^2m_h/2\approx0.002$ cm.

\section{Direct detection}
For direct detection of $MS\overline{M}$'s the interaction can be between the monopole and a nucleus or the string and a nucleus through Aharonov-Bohm scattering. The differential cross section for the classical scattering of an electric charge off the field of a magnetic monopole for large impact parameter is given by \cite{pietenpol}
\begin{equation}
\frac{d\sigma_{MN}}{d\Omega}=\bigg(\frac{eg}{E}\bigg)^2\frac{v^2}{\theta^4},
\end{equation}
where $E$ and $v$ are  the energy and speed of the incoming charge. The differential cross section (per unit length) for AB scattering of an electric charge off a string is given by
\begin{equation}
\frac{d\sigma_{AB}}{d\theta}=\frac{\sin^2(\pi\epsilon)}{2\pi p\sin^2(\theta/2)},
\end{equation}
where $p$ is the transverse particle momentum.

We can write the above expressions in terms of a nucleon recoil energy $E_R$, which is the physical quantity measured.  The results are
\begin{equation}
\frac{d\sigma_{MN}}{dE_R}=\frac{2\pi Z^2\alpha\alpha_M\chi^2}{m_NE_R^2}
\end{equation}
and
\begin{equation}
\frac{d\sigma_{AB}}{dE_R}\approx\frac{4L}{3}\frac{\pi\chi^2 v}{E_R^2}\bigg(\frac{2m_Nv^2}{E_R}-1\bigg)^{-1/2}.
\end{equation}
For the latter we account for two strings of length $L$, with each on average $2/3$ shorter due to the oscillations.

Among the direct detection experiments, the best upper bound on the interaction cross section for heavy dark matter (several hundred GeV to TeV masses) is given by the CDMS experiment \cite{cdms}. This experiment uses germanium or silicon crystals as the absorber and has a sensitivity threshold $E_R\approx 10$ keV. We use $m_N=26$ GeV and $Z=14$ corresponding to silicon, the lighter of the two nuclei.  The integrals that determine the total cross sections are dominated by the lower limit, the minimum recoil energies.  We find
\begin{eqnarray}
\sigma_{MN}&\approx& 10^{-54} \text{ cm}^2,\\
\sigma_{AB}&\approx& \bigg(\frac{L}{10^{-9}\mbox{ cm}}\bigg)10^{-50}\text{ cm}^2,
\end{eqnarray}
where we have set $v=10^{-3}$. Thus the Aharonov-Bohm scattering will be the dominant form of interaction. The current sensitivity is $\sigma_{exp}\approx10^{-43}(m_{DM}/1\mbox{ TeV})\mbox{ cm}^2$. Thus for a typical $L$ not much above $10^{-9}$ cm, the $AB$ cross section is still orders of magnitude below the best sensitivities available today.

For monopole-nucleus scattering we should comment on the maximum recoil energy, $E_{R{\rm max}}=2m_Nv^2$, since the $M$ and $\overline{M}$ can move at speeds $\sim(\mu L/m_M)^{1/2}$ which could be much greater than $10^{-3}c$. But the $E_{R{\rm max}}$ is limited by another effect. Recall that in the massless $\gamma_h$ limit the mutual coupling between the $M$'s and normal charges vanishes according to the Dirac quantization condition. Thus for distances of approach less than $m_h^{-1}$ the monopoles become invisible to normal charges and therefore scattering events where $q^2\gtrsim m_h^2$ cannot occur. This implies that $E_{R{\rm max}}\approx m_h^2/2m_N$, and at this energy a bump could be expected in the spectrum due to a type of pile-up effect.

\section{Outlook}
We have presented a dark matter candidate, the $MS\overline{M}$, with extremely weak couplings to the observable world (kinetic mixing parameter $\chi\approx10^{-15}$). When it decays the annihilating monopoles contribute to high energy electron/positron, neutrino and gamma ray signals. Excited strings and string loops may also be a source of low energy $e^+e^-$ pairs. The self-interactions of $MS\overline{M}$'s can have implications for dark halo dynamics. But there is little chance of observing $MS\overline{M}$'s through direct detection or in collider searches. The extremely weak coupling manifests itself as a long lifetime of the particle that mediates the interactions with the dark sector, the $\gamma_h$. Only $\gamma_h$'s that originate farther from us than its typical travel distance will be observable. This travel distance can be somewhat larger than the distance to the galactic core and this could reduce the anisotropy of the gamma ray and neutrino signals.

Meanwhile both the monopole mass and the $\gamma_h$ mass will be constrained by the characteristics of the observed $e^\pm$ spectra. If $m_h$ is indeed significantly larger than 1 MeV then it will produce an apparent puzzle given the constraints from BBN. This in turn would provide indirect evidence for strong interactions in the hidden sector since, as we have described, in this case energy is dumped into relativistic cosmic strings rather than nonrelativistic $\gamma_h$'s. A determination of the $\gamma_h$ travel distance would also fix the mixing parameter $\chi$ through (\ref{e18}). The parameters of the model are then determined. We may then find that a ``miracle'' has occurred, since it could turn out that the model gives both the right abundance and the right lifetime for the $MS\overline{M}$'s.

\section*{Acknowledgments}
We thank M. Pospelov for a useful discussion, T. Vachaspati for correspondence and W. Cheung and S. Freedman for their comments. This work was supported in part by the Natural Science and Engineering Research Council of Canada.


\begin{thebibliography}{99}
\bibitem{olum} J. J. Blanco-Pillado and K. D. Olum, Phys. Rev. D60, 083001 (1999), astro-ph/9904315.
\bibitem{parker} E. N. Parker, Astr. J. {160}, 383 (1970).
\bibitem{vilenkin} A. Vilenkin, Nucl. Phys.  {B196}, 240 (1982).
\bibitem{vil1} T. Vachaspati and A. Vilenkin, Phys. Rev. D35, 1131 (1987).
\bibitem{vil2} V. Berezinsky, X. Martin and A. Vilenkin, Phys. Rev. D56, 2024 (1997).
\bibitem{olum1} J. J. Blanco-Pillado and K. D. Olum, JCAP 1005, 014 (2010), arXiv:0707.3460.
\bibitem{vash1} T. Vachaspati, Phys. Rev. D80, 063502 (2009), arXiv:0902.1764.
\bibitem{pamela} O. Adriani et al., Nature 458, 607 (2009), arXiv:0810.4995.
\bibitem{fermi} A. A. Abdo et al., Phys. Rev. Lett. 102, 181101 (2009), arXiv:0905.0025.
\bibitem{holdom} B. Holdom, Phys. Lett. {166B}, 196 (1986).
\bibitem{nima} N. Arkani-Hamed, D. P. Finkbeiner, T. R. Slatyer, N. Weiner, Phys. Rev. D79, 015014 (2009), arXiv:0810.0713.
\bibitem{pospelov}M. Pospelov and A. Ritz, Phys. Lett. B671, 391 (2009), arXiv:0810.1502.
\bibitem{vilshell} A. Vilenkin and E. P. S. Shellard,``Cosmic strings and other topological defects",  Cambridge University Press, Cambridge, 2000.
\bibitem{meade} P. Meade, M. Papucci, A. Strumia, T. Volansky, Nucl. Phys. B831, 178 (2010), arXiv:0905.0480; M. Papucci and A. Strumia, JCAP 1003, 014 (2010) arXiv:0912.0742.
\bibitem{posp} M. Pospelov, A. Ritz, M. B. Voloshin, Phys. Lett. B662, 53 (2008), arXiv:0711.4866.
\bibitem{jaeckel} F. Br\"ummer and J. Jaeckel, Phys. Lett. B675, 360 (2009), arXiv:0902.3615.
\bibitem{preskill} J. Preskill, Phys. Rev. Lett. {43}, 1365 (1979).
\bibitem{kibble} T.W.B. Kibble, J. Phys. A {9}, 1387 (1976).
\bibitem{zurek} W. H. Zurek, Nature {317}, 505 (1985).
\bibitem{murayama} H. Murayama and J. Shu, Phys. Lett. B686, 162 (2010), arXiv:0905.1720.
\bibitem{kolb} E.W. Kolb and M. S. Turner, The early universe (Frontiers in Physics, Reading, MA: Addison-Wesley, 1990).
\bibitem{cohen}A.G. Cohen, D.B. Kaplan, A.E. Nelson, Phys. Lett. B412, 301 (1997), arXiv:hep-ph/9706275.
\bibitem{hill} C. Hill, Nucl. Phys. B224, 469 (1983).
\bibitem{rothstein} I.Z. Rothstein, T. Schwetz, J. Zupan, JCAP 0907, 018 (2009), arXiv:0903.3116.
\bibitem{ibarra} A. Ibarra, D. Tran, and C. Weniger, Phys. Rev. D81, 023529 (2010), arXiv:0909.3514.
\bibitem{profumo} L. Dugger, T.E. Jeltema, S. Profumo, JCAP 12, 015 (2009), arXiv:1009.5988.
\bibitem{casper} P. Casper, B. Allen, Phys. Rev. D52, 4337 (1995), gr-qc/9505018.
\bibitem{martin} X. Martin, A. Vilenkin, Phys. Rev. D55, 6054 (1997), arXiv:gr-qc/9612008.
\bibitem{alford} R. Rohm, Ph.D. Thesis, Princeton University, 1985; M.G. Alford and F. Wilczek, Phys. Rev. Lett. 62, 1071 (1989).
\bibitem{garf} D. Garfinkle and T. Vachaspati, Phys. Rev. D36, 2229 (1987).
\bibitem{vash4} Y-Z. Chu, H. Mathur, T. Vachaspati, Phys. Rev. D82, 063515  (2010), arXiv:1003.0674.
\bibitem{garriga} J. Garriga, D. Harari and E. Verdaguer, Nucl. Phys. B339, 560 (1990).
\bibitem{vash2} K. Jones-Smith, H. Mathur, T. Vachaspati, Phys. Rev. D81, 043503 (2010), arXiv:0911.0682.
\bibitem{vash3} D.A. Steer and T. Vachaspati, Phys. Rev. D83, 043528 (2011), arXiv:1012.1998.
\bibitem{everett} A.E. Everett, Phys. Rev. {D24}, 858 (1981).
\bibitem{nielsen} H.B. Nielsen and P. Olesen, Nucl. Phys. B61, 45 (1973).
\bibitem{jackiw} R. Jackiw and R. Rossi, Nucl. Phys. B190, 681 (1980).
\bibitem{bbn} Y.I. Izotov and T.X. Thuan, Astrophys. J. 710, L67 (2010), arXiv:1001.4440; E. Aver, K.A. Olive, and E.D. Skillman, JCAP 1005, 003 (2010), arXiv:1001.5218.
\bibitem{cmb} E. Komatsu et al., Astrophys. J. Suppl. 192, 18 (2011), arXiv:1001.4538; J. Dunkley et al. (2010), arXiv:1009.0866.
\bibitem{press} C.J. Copi and T. Vachaspati, Phys. Rev. D83, 023529 (2011), arXiv:1010.4030; R. J. Scherrer and W.H. Press, Phys. Rev. D39, 371 (1989).
\bibitem{spergel} D.N. Spergel and P.J. Steinhardt, Phys. Rev. Lett. 84, 3760 (2000).
\bibitem{jean} P. Jean et al., Astron. Astrophys. 407, L55 (2003); J. Knodlseder et al., Astron. Astrophys. 411, L457 (2003), arXiv:astro-ph/0309442.
\bibitem{maxim} C. Picciotto and M. Pospelov, Phys. Lett. B605, 15 (2005), arXiv:hep-ph/0402178.
\bibitem{pietenpol} I.R. Lapidus and J.L. Pietenpol, Am. J. Phys. {28}, 17 (1960).
\bibitem{cdms} Z. Ahmed et al., Science 327, 1619 (2010), arXiv:0912.3592.

\end{thebibliography}
\end{document}